\documentclass[aps,pra,reprint,superscriptaddress,amsmath,amssymb,floatfix]{revtex4-2}

\usepackage{graphicx}% Include figure files
\usepackage{dcolumn}% Align table columns on decimal point
\usepackage{bm}% bold math
\usepackage{xcolor}
\usepackage{soul}
\usepackage{xfrac}

\usepackage[colorlinks=true,bookmarks=false,allcolors=blue,citecolor=blue,urlcolor=blue]{hyperref} %latex w/dvips

\DeclareGraphicsExtensions{.pdf,.jpg, .png}
\usepackage{epstopdf}
\usepackage{enumitem}

\usepackage{cleveref}

\begin{document}

\title{Fourier transform of the hyperbola and its role in hyperbolic photonics}

\author{Emroz Khan}
\email{ekhan@gc.cuny.edu} % Un-comment this for the final version!
\affiliation{Photonics Initiative, Advanced Science Research Center, The City University of New York, New York,  NY 10031, USA}

\author{Andrea Al\`u}
\affiliation{Photonics Initiative, Advanced Science Research Center, The City University of New York, New York,  NY 10031, USA}
\affiliation{Physics Program, Graduate Center of the City University of New York, New York, NY 10016, USA}

%\address{$^1$Photonics Initiative, Advanced Science Research Center, The City University of New York, New York,  NY 10031, USA \\
%$^2$Physics Program, Graduate Center of the City University of New York, New York, NY 10016, USA}

\begin{abstract}
Motivated by recent breakthrough studies of wave hyperbolicity in extremely anisotropic natural materials and artificial composites, we investigate the radiation pattern of a localized emitter in a hyperbolic medium. Since the emission of a point source is associated with the Fourier transform of the iso-frequency contours of a medium, we derive and analyze the properties of the Fourier transform of hyperbolic dispersion, which sheds light into the emission properties in the presence of hyperbolic bands. Our analysis leads to a generalized form of Huygens’ principle for hyperbolic waves, connecting to the emergence of negative refraction and focusing with hyperbolic media. We also highlight the occurrence of aliasing artifacts in polariton imaging. More broadly, our findings provide analytical tools to model polariton propagation in materials with extreme anisotropy, and may be applied to several other physical platforms featuring hyperbolic responses, from astrophysics to seismology.
\end{abstract}

\maketitle

\section{Introduction}
\label{Sec:I}
Hyperbolas arise naturally in the description of different phenomena. Comets that pass through our solar system only once using the Sun as a gravitational slingshot trace out hyperbolic trajectories~\cite{hughes1991hyperbolic, sky}. So do the alpha particles scattered by the Coulomb potential of atomic nuclei in a Rutherford experiment~\cite{rutherford}. In optics, the interference pattern produced by two localized sources shows hyperbolic fringes on a screen that is placed in parallel to the line connecting the two sources~\cite{steel1983interferometry}. Similarly, the shadow-tip of a pole can run along a hyperbola on the ground over the course of a day, and as a result, hyperbolas are seen as declination lines on a sundial~\cite{rohr2012sundials}. One can in fact notice hyperbolic shadow on their own bedroom wall emanating from an adjacent lamp shade~\cite{horst2001shape}. Hyperbolicity can also enter the abstract mathematical description of numerous physical systems. For example, in  reflection seismology, the relationship between the arrival time of the reflected signal and the source-to-receiver distance, known as the offset, is hyperbolic~\cite{sheriff1995exploration}. In radio navigation systems like Gee or LORAN, hyperbolas are the basis for locating a point from the differences in its distances to a given set of points~\cite{navigation}. In fact, any inverse relationship between two physical quantities, for example, Boyle's law or Wien's law, can be associated with a rectangular hyperbola. \\

\begin{figure}[t]
\centering
	\includegraphics[width=\columnwidth]{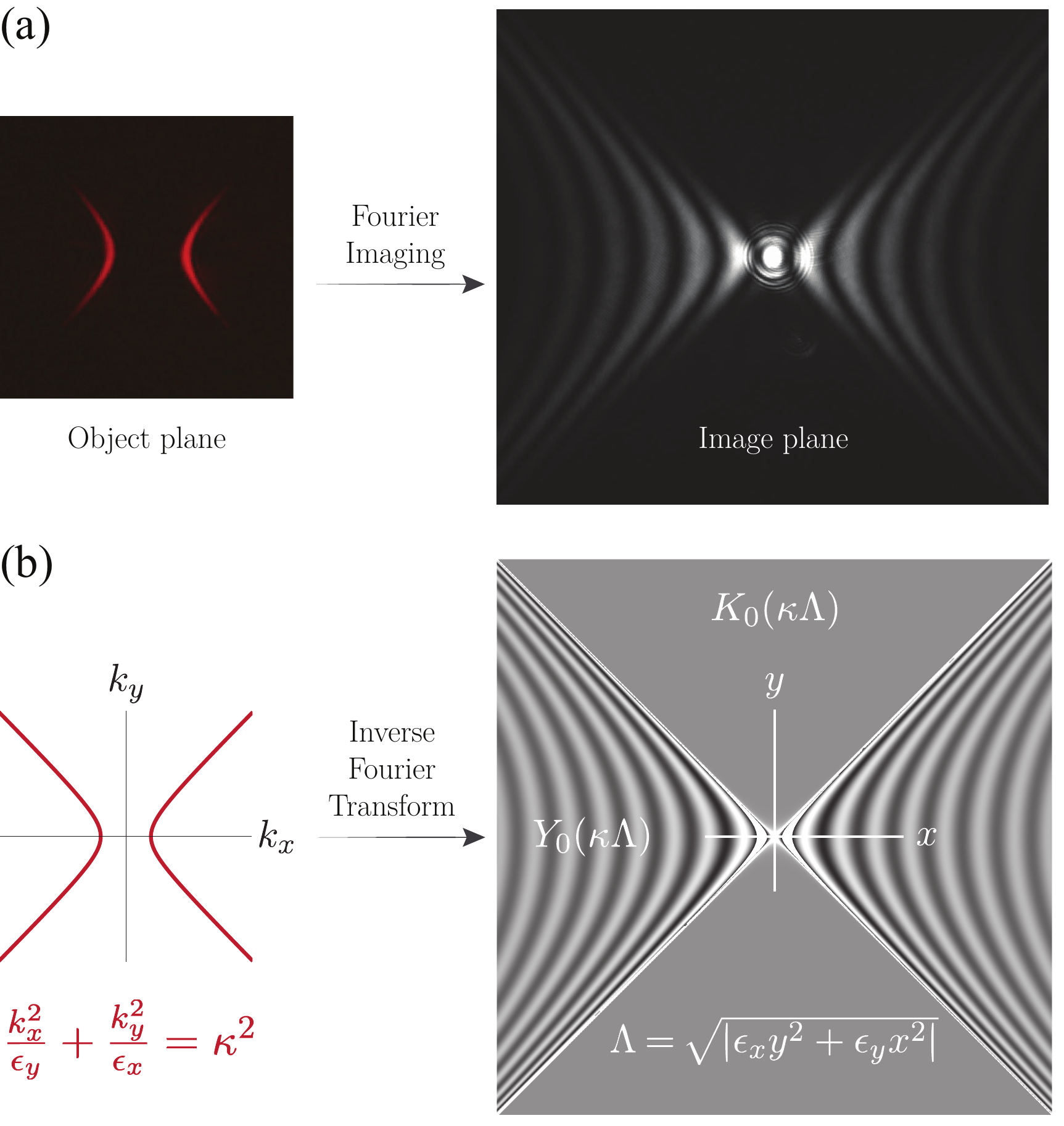}
\caption{Fourier transform of a hyperbola obtained (a)~experimentally via Fourier imaging of a pair of truncated hyperbolic slits;  (b)~analytically for an infinite hyperbola using our formulation. In both cases, rectangular hyperbolas with $\epsilon_y = -\epsilon_x = 1$ were considered (left panels), and the corresponding Fourier patterns exhibit distinct hyperbolic fringes (right panels).}
\label{Fig:intro}
\end{figure}

In modern material science and photonics, hyperbolas have been playing an important role. When electromagnetic materials (or metamaterials) display extremely anisotropic responses, with oppositely-signed real parts of permittivity components for two orthogonal directions, they lead to a hyperbolic dispersion of the iso-frequency contours of their eigenmodes~\cite{jacob2006optical,salandrino2006far,liu2007far,esslinger2014tetradymites,dai2014tunable,li2009experimental,quan2019hyperbolic}, either in their bulk or at their interface with air. This leads to exciting opportunities for nanoscience, with directional and highly confined optical beams supported by such \textit{hyperbolic materials}. Their features can be accessed by exciting hyperbolic media with localized sources. Generally, localized excitations of a medium access all available momenta, leading to an emission pattern that is the spatial Fourier transform of the iso-frequency dispersion contour. For isotropic media, a localized emitter generates a field profile that follows a bullseye pattern, which is indeed the Fourier transform of a circular iso-frequency dispersion curve~\cite{lezec2002beaming}. Similarly, if a material supports a hyperbolic iso-frequency contour at a given frequency, the emission from a localized emitter will feature a field distribution associated with the Fourier transform of a hyperbola. In order to image its pattern, we can rely on near-field scanning optical experiments, as done in many recent papers~\cite{ma2021ghost,passler2022hyperbolic,hu2023real,ni2023observation}, or rely on another platform to realize optically the Fourier transform of a spatial pattern, i.e., using an optical lens~\cite{goodman2005introduction}.  Figure~\ref{Fig:intro}a shows the experimental image plane distribution created by a lens, while imaging two slits with hyperbolic profile in the object plane. In the following, we analytically evaluate this field pattern, and apply this result to provide physical insights into wave propagation in hyperbolic media and other phenomena featuring the Fourier transform of a hyperbolic curve. \\

The analytical evaluation of the Fourier transform of an hyperbola poses some challenges, because of its unbounded support. Because of Parseval’s theorem~\cite{soliman1990continuous}, this implies that we can expect the Fourier transform to show  singularities. For a hyperbolic iso-frequency dispersion characterized by
\begin{equation}
\frac{k_x^2}{\epsilon_y}+\frac{k_y^2}{\epsilon_x}=\kappa^2
 \label{Eq:hyperbola}
\end{equation}
in the reciprocal space with spatial frequency (also referred to as wavevector or momentum) components $k_x$ and $k_y$, the inverse Fourier transform $f(x,y)$ in real space is
\begin{equation}
f(x,y)=\begin{cases}
  -2\pi \kappa \sqrt{|\epsilon_x \epsilon_y|}\: Y_0(\kappa \Lambda) & \text{major region}\\      
  4 \kappa \sqrt{|\epsilon_x \epsilon_y|}\: K_0(\kappa \Lambda) & \text{minor region}\\      
\infty & \text{separatrix}\\      
\end{cases}
 \label{Eq:hypFT}
\end{equation}
where  
\begin{equation}
\Lambda = \sqrt{|\epsilon_x y^2 + \epsilon_y x^2|}
\label{Eq:hyperadius}
\end{equation}
denotes a ``hyper-radius" coordinate, and $Y_0$ and $K_0$ are the Bessel and modified Bessel functions, respectively, both of the second kind and order zero~\cite{arfken2011mathematical}. Here the three regions are defined as (see Fig.~\ref{Fig:intro}b for the case $\epsilon_x<0$ and $\epsilon_y>0$)
\begin{equation}
\begin{aligned}
\text{major region:} &\quad \epsilon_x y^2 + \epsilon_y x^2 > 0\\
\text{minor region:} &\quad \epsilon_x y^2 + \epsilon_y x^2 < 0\\
\text{separatrix:} &\quad \epsilon_x y^2 + \epsilon_y x^2 = 0.
\end{aligned}
\label{Eq:region}
\end{equation}

In the following sections we derive and interpret this result. Section~\ref{Sec:M} introduces the method employed for evaluating the Fourier transform and derives the analytical result in~(\ref{Eq:hypFT}). Section~\ref{Sec:GR} graphically analyzes the obtained Fourier pattern, while Section~\ref{Sec:PI} provides a physical interpretation in terms of an ensemble of linear fringes. Based on these findings, Section~\ref{Sec:D}  generalizes Huygens’ principle for waves in  hyperbolic media and Section~\ref{Sec:A}  explains commonly encountered image aliasing artifacts.

\section{Analytical Derivation}
\label{Sec:M}
If we consider the known case of a circle of  radius $\kappa_0$, characterized by the dispersion in reciprocal space
\begin{equation}
F(k) = \delta(k-\kappa_0),
\label{Eq:circleF}
\end{equation}
a straightforward application of the Hankel transform~\cite{arfken2011mathematical} gives us its inverse Fourier transform as 
\begin{equation}
f(r) = 2 \pi \kappa_0\: J_0(\kappa_0 r)
\label{Eq:circle}
\end{equation}
where $J_0$ is the Bessel function of the first kind of order zero. Here $k$ and $r$ are the radial coordinates in reciprocal and real spaces, respectively. As the Hankel transform is only applicable for radially symmetric functions, this method cannot be extended to hyperbolic curves. Also, since a circle can not be continuously deformed into a hyperbola in the topological sense, one curve being closed while the other being open, it is not straightforward either to arrive at the Fourier transform of a hyperbola from the one of a circle, as it would be for an ellipse, for example. \\

With this caveat, we turn to the original definition of inverse Fourier transform of a two-dimensional function $F(k_x,k_y)$ in  reciprocal space, namely
\begin{equation}
f(x,y) = \int_{k_x=-\infty}^{k_x=\infty}\int_{k_y=-\infty}^{k_y=\infty} F(k_x,k_y) e^{i(k_x x + k_y y)} \:dk_x \:dk_y.
\label{Eq:def}
\end{equation}
For the previous case of a circle, Eq.~(\ref{Eq:def}) becomes
\begin{widetext}
\begin{equation}
\begin{aligned}
f(x,y)& = \int_{k_y=-\infty}^{k_y=\infty} \left[  \int_{k_x=-\infty}^{k_x=\infty} \delta\left(\sqrt{k_x^2+k_y^2}-\kappa_0\right)  e^{i(k_x x + k_y y)} \:dk_x \right] \:dk_y\\
 & = \int_{k_y=-\kappa_0}^{k_y=\kappa_0} \left[  \int_{k_x=-\infty}^{k_x=\infty} \sum_{\pm}
 \delta\left(k_x \mp \sqrt{\kappa_0^2-k_y^2} \right) \frac{\kappa_0}{ \sqrt{\kappa_0^2-k_y^2} }  e^{i(k_x x + k_y y)} \:dk_x \right] \:dk_y\\
 & = \kappa_0 \sum_{\pm} \int_{k_y=-\kappa_0}^{k_y=\kappa_0}   \frac{e^{i\left(k_y y \: \pm  \sqrt{\kappa_0^2-k_y^2} x \right)}}{ \sqrt{\kappa_0^2-k_y^2} } \:dk_y.
 \end{aligned}
\label{Eq:defc}
\end{equation}
\end{widetext}
In the intermediate step here, we have used the well-known~\cite{arfken2011mathematical} delta function identity 	$\delta\left( g(x) \right) = \sum_i {\delta(x-x_i)}/{\left| g'(x_i) \right|}$ with $x_i$'s being real roots of $g(x)=0$. With the substitution $k_y = \kappa_0 \sin\phi$ and introducing the polar angle coordinate $\theta=\arctan(y/x)$ in real space, we obtain
%\begin{widetext}
\begin{equation}
\begin{aligned}
f(r,\theta) &  = \kappa_0 \sum_{\pm} \int_{\phi=-\pi/2\, \mp \, \theta}^{\phi=\pi/2\, \mp\, \theta} e^{\pm i \kappa_0 r \cos\phi}  \:d\phi\\
& = \kappa_0 \left[ \int_{\phi=-\pi/2\, - \, \theta}^{\phi=\pi/2\, - \, \theta} 
+ 					\int_{\phi=\pi/2\, + \, \theta}^{\phi=-\pi/2\, + \, \theta} \right] 
e^{ i \kappa_0 r \cos\phi}  \:d\phi.
 \end{aligned}
 \label{Eq:limsC}
\end{equation}
%\end{widetext}
%Now it is easy to check that the integrand takes on the same values for the intervals	$\pi/2 \leq  \phi \leq \pi/2 + \theta$ and  $-\pi/2 \geq  \phi \geq -(\pi/2 + \theta)$, as well as  for the two intervals $\pi/2 - \theta \leq  \pm\phi \leq \pi/2$.
Now for the second integral a final change in variable $\phi \rightarrow -\phi $ gives us a combined integration limit spanning over all angles, resulting in
\begin{equation}
f(r,\theta)   =   \kappa_0 \int_{\phi=-\pi}^{\phi=\pi}  e^{ i \kappa_0 r \cos\phi}  \:d\phi
 \label{Eq:J0}
\end{equation} 
from which~(\ref{Eq:circle}) immediately follows~\cite{arfken2011mathematical}. \\

Note that in this method the key step was to split the original delta function in its argument's roots and integrate them out by one of the reciprocal coordinates. This method also works for the known case of a straight line characterized by $k_y = \mu k_x$. In this case, the inverse Fourier transform becomes
\begin{equation}
\begin{aligned}
f(x,y) &  = \iint_{k_x, k_y=-\infty }^{k_x,k_y=\infty}  \delta\left( k_y - \mu k_x  \right)   e^{i(k_x x + k_y y)} \:dk_x \:dk_y\\
& =\int_{k_x=-\infty}^{k_x=\infty}   e^{ik_x(x +\mu y )  } \:dk_x\\
& = 2 \pi\:  \delta(x+\mu y)
\end{aligned}
 \label{Eq:line}
\end{equation}
which characterizes a perpendicular  line in real space. Here, of course, we assume the conventional choice of aligning the coordinate axes with their reciprocal counterparts. The orthogonality nature of the above result can be understood in the following way: a straight line stretches infinitely in one direction while in the orthogonal direction, it has a ``zero" width given by a $\delta$-spike. Therefore, based on the uncertainty relation of Fourier analysis, the corresponding transform should have a sharp $\delta$-like feature and an infinite support in these two directions, respectively, which results in a straight line perpendicular to the original one. Also, note how the use of $\delta$-functions can dramatically ease the Fourier calculations for functions with infinite extent.\\

We now apply this method to a hyperbola in reciprocal space characterized by~(\ref{Eq:hyperbola}), that is,
\begin{equation}
F(k_x, k_y) = \delta\left( \sqrt{\frac{k_x^2}{\epsilon_y}+\frac{k_y^2}{\epsilon_x}}-\kappa \right)
 \label{Eq:hypFD}
\end{equation}
where $\epsilon_x \epsilon_y<0$. Without loss of generality, we assume $\epsilon_x<0$ and $\epsilon_y>0$. Splitting the $\delta$-function in its $k_x$ argument and carrying out the $k_x$-integral in its inverse Fourier transform, we get
\begin{equation}
f(x,y) = I_++I_-
 \label{Eq:hypF}
\end{equation}
where 
\begin{equation}
I_{\pm} =  \int_{k_y=-\infty}^{k_y=\infty}   \frac{\kappa \epsilon_y}{ \sqrt{\epsilon_y\kappa^2-\frac{\epsilon_y}{\epsilon_x}k_y^2} }\: e^{i\left(k_y y \: \pm  \sqrt{\epsilon_y\kappa^2-\frac{\epsilon_y}{\epsilon_x}k_y^2} x \right)}  \:dk_y
 \label{Eq:ipm}
\end{equation}
which, upon the substitution $k_y = \sqrt{-\epsilon_x} \, \kappa \sinh \alpha$,   becomes
\begin{equation}
I_{\pm} = \kappa \sqrt{|\epsilon_x \epsilon_y|} \int_{\alpha=-\infty}^{\alpha=\infty}  e^{i \kappa \left(  \sqrt{-\epsilon_x}\, y \sinh \alpha \: \pm \: \sqrt{\epsilon_y} \, x  \cosh \alpha \right) }    \: d\alpha.
 \label{Eq:sinh}
\end{equation}

Since a hyperbola is invariant under reflections about its major and minor axes, i.e.,   $F(k_x, k_y)=F(|k_x|,  |k_y|)$, its  transform will also respect this symmetry by maintaining $f(x,y) = f(|x|,|y|)$. Hence it suffices to restrict our attention to  $x,y\geq 0$. Depending on the sign of $\epsilon_x y^2 + \epsilon_y x^2$, we can partition  the real space into three regions as defined in~(\ref{Eq:region}). We evaluate~(\ref{Eq:sinh}) in these regions case by case below.\\

\noindent
\textit{Major region: } $\epsilon_x y^2 + \epsilon_y x^2>0$\\
In terms of the hyper-radius $\Lambda$ as defined in~(\ref{Eq:hyperadius}), let $\sinh \beta = \sqrt{-\epsilon_x}\, y/\Lambda$ so that~(\ref{Eq:sinh}) reduces to 
\begin{equation}
\begin{aligned}
I_{\pm} &= \kappa \sqrt{|\epsilon_x \epsilon_y|} 
\int_{\alpha=-\infty}^{\alpha=\infty}
  e^{\pm i \kappa \Lambda   \cosh \alpha  }    \: d\alpha\\
  & =  - \pi \kappa \sqrt{|\epsilon_x \epsilon_y|} \left( Y_0(\kappa \Lambda) \mp i J_0(\kappa \Lambda)  \right) 
  \end{aligned}
 \label{Eq:cal1}
\end{equation}
where we have employed the Mehler-Sonine integrals in the last step~\cite{gradstein2007table}. Then from~(\ref{Eq:hypF}) we have
\begin{equation}
f(x,y) =  - 2 \pi \kappa \sqrt{|\epsilon_x \epsilon_y|} \: Y_0(\kappa \Lambda)  \:\: \text{for} \:\: \epsilon_x y^2 + \epsilon_y x^2>0.
 \label{Eq:reg1}
\end{equation}
\\
\noindent
\textit{Minor region: } $\epsilon_x y^2 + \epsilon_y x^2<0$\\
Again in terms of the hyper-radius, if we let $\sinh \gamma = \sqrt{\epsilon_y}\, x/\Lambda$,~(\ref{Eq:sinh}) yields to
\begin{equation}
I_{\pm} = \kappa \sqrt{|\epsilon_x \epsilon_y|} 
\int_{\alpha=-\infty}^{\alpha=\infty}
  e^{ i \kappa \Lambda   \sinh \alpha  }    \: d\alpha
 \label{Eq:cal2}
\end{equation}
which, through another Mehler-Sonine integral~\cite{gradstein2007table}, gives us the final result
\begin{equation}
f(x,y) =   4 \kappa \sqrt{|\epsilon_x \epsilon_y|} \: K_0(\kappa \Lambda)  \:\: \text{for} \:\: \epsilon_x y^2 + \epsilon_y x^2<0.
 \label{Eq:reg2}
\end{equation}
\\
\noindent
\textit{Separatrix: } $\epsilon_x y^2 + \epsilon_y x^2=0$\\
In this case, let $\eta = \sqrt{\epsilon_y}\, x = \sqrt{-\epsilon_x}\, y  $, so that~(\ref{Eq:sinh}) becomes
\begin{equation}
I_{\pm} = \kappa \sqrt{|\epsilon_x \epsilon_y|} 
\int_{\alpha=-\infty}^{\alpha=\infty}
  e^{ \pm i \kappa \eta \, e^{\pm\alpha}  }    \: d\alpha.
 \label{Eq:cal3}
\end{equation}
To calculate $I_+$ we make the substitution $t = e^\alpha$ which gives us
\begin{equation}
\begin{aligned}
\frac{I_{+}}  {\kappa \sqrt{|\epsilon_x \epsilon_y|} }=
\int_{t=0}^{t=\infty}
  \frac{e^{  i \kappa \eta \, t }}{t}    \: dt 
 & =  \text{Ei}(i\infty)-\text{Ei}(0)\\
 & =\infty+i\pi
  \end{aligned}
 \label{Eq:cal3p}
\end{equation}
where $\text{Ei}$ is the exponential integral~\cite{gradstein2007table}. Similarly $\tfrac{I_-}{ \kappa \sqrt{|\epsilon_x \epsilon_y|}}$ evaluates to $\infty-i\pi$ making
\begin{equation}
f(x,y) =   \infty  \:\: \text{for} \:\: \epsilon_x y^2 + \epsilon_y x^2=0.
 \label{Eq:reg3}
\end{equation}
\\

Combining the results for the three regions from~(\ref{Eq:reg1}), (\ref{Eq:reg2})~and~(\ref{Eq:reg3}), and using the reflection property of the hyperbola, we finally have its complete inverse Fourier transform for all points in the real space,  as   in~(\ref{Eq:hypFT})--(\ref{Eq:region}). Although the algebraic expressions were derived for $\epsilon_x<0$ and $\epsilon_y>0$, they remain invariant under a sign reversal of $\epsilon_x$ and $\epsilon_y$. 
Note that  the transform is real-valued where it is bounded, as expected.
Now with all the algebra completed, we analyze the transform expressions   in the next section.  

\section{Graphical representation}
\label{Sec:GR}

In order to gain a pictorial understanding of the inverse Fourier transform  of a hyperbola, as characterized  in~(\ref{Eq:hyperbola})~and~(\ref{Eq:hypFT}), we analyze the transform in the three regions  of real space  defined in~(\ref{Eq:region}). \\

\begin{figure}[t]
\centering
	\includegraphics[width=\columnwidth]{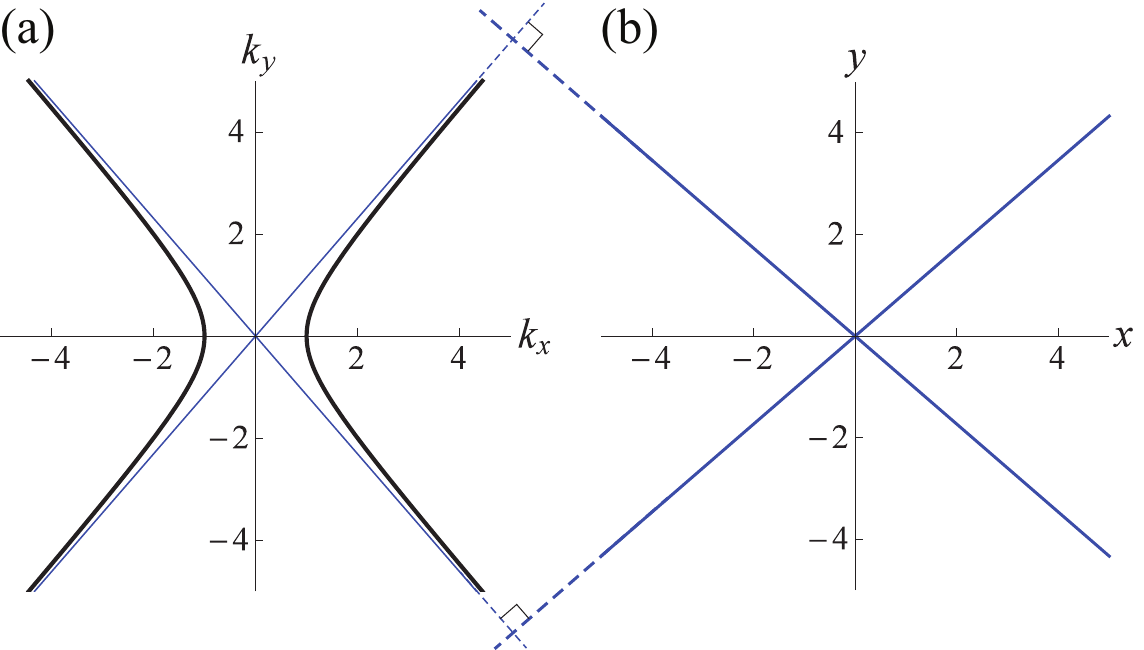}
\caption{Fourier transform of a hyperbola in the separatrix is characterized by a pair of straight lines where its functional value is unbounded. For a given hyperbola in reciprocal space~(a), each of its two asymptotes is perpendicular to one of the separatrix lines in the corresponding transform~(b). This orthogonality relation is a natural consequence of the transform holding true in the limiting case of $\kappa \rightarrow 0$.}
\label{Fig:regIII}
\end{figure}

\noindent
\textit{Separatrix: } $\epsilon_x y^2 + \epsilon_y x^2=0$\\
We start with this simplest region, which is characterized by a pair of straight lines given by $y = \pm \sqrt{-\sfrac{\epsilon_y}{\epsilon_x}}\,x$. The functional value of the transform for all points in this region is unbounded. This ``blow-up" is acceptable since our hyperbola has an infinite support. Note that, as the two asymptotes of the given hyperbola in reciprocal space are described by $k_y = \pm \sqrt{-\sfrac{\epsilon_x}{\epsilon_y}}\,k_x$ (see Fig.~\ref{Fig:regIII}a which depicts the case of $\epsilon_x<0$ and $\epsilon_y>0$), each of the asymptotes is normal to one of the separatrix lines, as shown in Fig.~\ref{Fig:regIII}b. This orthogonality relation can be understood in the limit of $\kappa \rightarrow 0$ where our hyperbola evolves into the degenerate case of a pair of straight lines. In this limiting case, the corresponding transform (see Eq.~(\ref{Eq:hypFT})) becomes zero in the major and minor regions, and blows up in the separatrix, giving us a pair of straight lines orthogonal to the original degenerate pair, which is expected based on our discussion of the Fourier transform of a straight line in Section~\ref{Sec:M}.
% Also note that  reversing the signs of $\epsilon_x$ and $\epsilon_y$ does not alter the description of this region which can also be understood in the $\kappa \rightarrow 0$ limit.\\
Moreover, this $\kappa \rightarrow 0$ picture also explains why the description of the separatrix is unaltered if the signs of $\epsilon_x$ and $\epsilon_y$ are reversed.
In fact, as the description does not depend on the value of $\kappa$ at all, and hence where the two foci of the hyperbola are located, one can conclude that the separatrix captures the Fourier transform information of the asymptotic behavior of a hyperbola.\\

\noindent
\textit{Minor region: } $\epsilon_x y^2 + \epsilon_y x^2<0$\\
As Fig.~\ref{Fig:regIII}b shows, the two intersecting lines of the separatrix partition the real plane into four ``quadrants" where $\epsilon_x y^2 + \epsilon_y x^2 \neq 0$. Among them,  those two quadrants that contain the reciprocal (conjugate) of the minor axis of a given hyperbola belong to the minor region. For the hyperbola illustrated in Fig.~\ref{Fig:regIII}a with its minor axis oriented along the $k_y$ axis, the corresponding minor region encompasses the upper and lower quadrants in Fig.~\ref{Fig:regIII}b.\\

\begin{figure}[tb]
\centering
	\includegraphics[width=0.8\columnwidth]{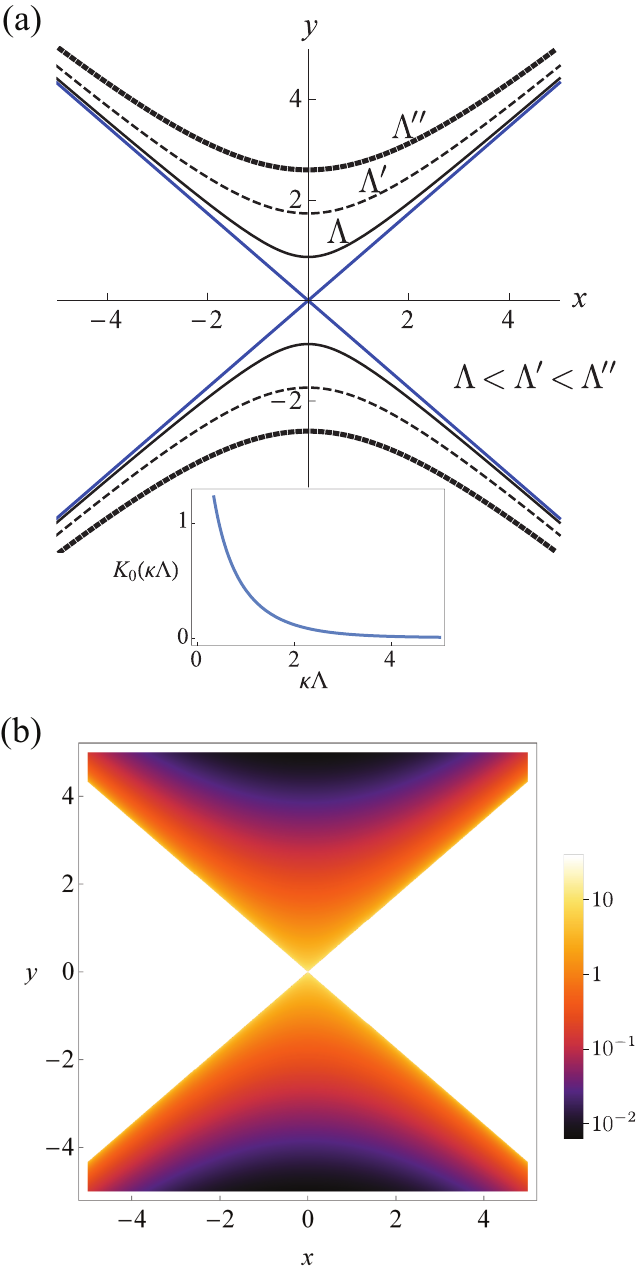}
\caption{Fourier transform of a hyperbola in the minor region shows sharp decay from the separatrix lines with hyperbolic contours. For the hyperbola in Fig.~\ref{Fig:regIII}a, this region contains the upper and lower quadrants of Fig.~\ref{Fig:regIII}b. The transform contours are given by a family of coasymptotic hyperbolas with three such hyperbolas shown in~(a) for different values of the hyper-radius $\Lambda, \Lambda'$ and $\Lambda''$. As the transform value decays quickly with the hyper-radius (inset of~(a)), we have the resulting transform in this region~(b) exhibit sharp decay with hyperbolic contours.
}
\label{Fig:regII}
\end{figure}

Since the functional value of the transform depends on the real coordinates $x$ and $y$ through the hyper-radius $\Lambda$, the transform contours are given by the constant-$\Lambda$ curves. In particular, for the minor region of our hyperbola in Fig.~\ref{Fig:regIII}, we obtain the contours from~(\ref{Eq:hyperadius}) as
\begin{equation}
\frac{y^2}{\Lambda^2/(-\epsilon_x)} - \frac{x^2}{\Lambda^2/\epsilon_y} = 1
 \label{Eq:contourII}
\end{equation}
which describes a family of hyperbolas parametrized by $\Lambda$ (see Fig.~\ref{Fig:regII}a), all with their major axes oriented along the $y$ direction. Although these hyperbolas are not confocal with their focal distance increasing linearly with the hyper-radius, they all, however, share the same asymptotes which also align with the straight lines of the separatrix. With the functional dependence of the transform on the hyper-radius shown in the inset of Fig.~\ref{Fig:regII}a, these ``coasymptotic" hyperbolas trace out different contour lines in the minor region (see Fig.~\ref{Fig:regII}b). Note how the transform shows sharp monotonic decay from the delta-like ridges of the separatrix with hyperbolic contours.\\

\noindent
\textit{Major region: } $\epsilon_x y^2 + \epsilon_y x^2>0$\\
Lastly, the major region encompasses those two quadrants that contain the reciprocal (conjugate) of the major axis of the hyperbola. For our hyperbola from Fig.~\ref{Fig:regIII}a, the remaining left and right quadrants in Fig.~\ref{Fig:regIII}b belong to this region. \\

\begin{figure}[tb]
\centering
	\includegraphics[width=0.8\columnwidth]{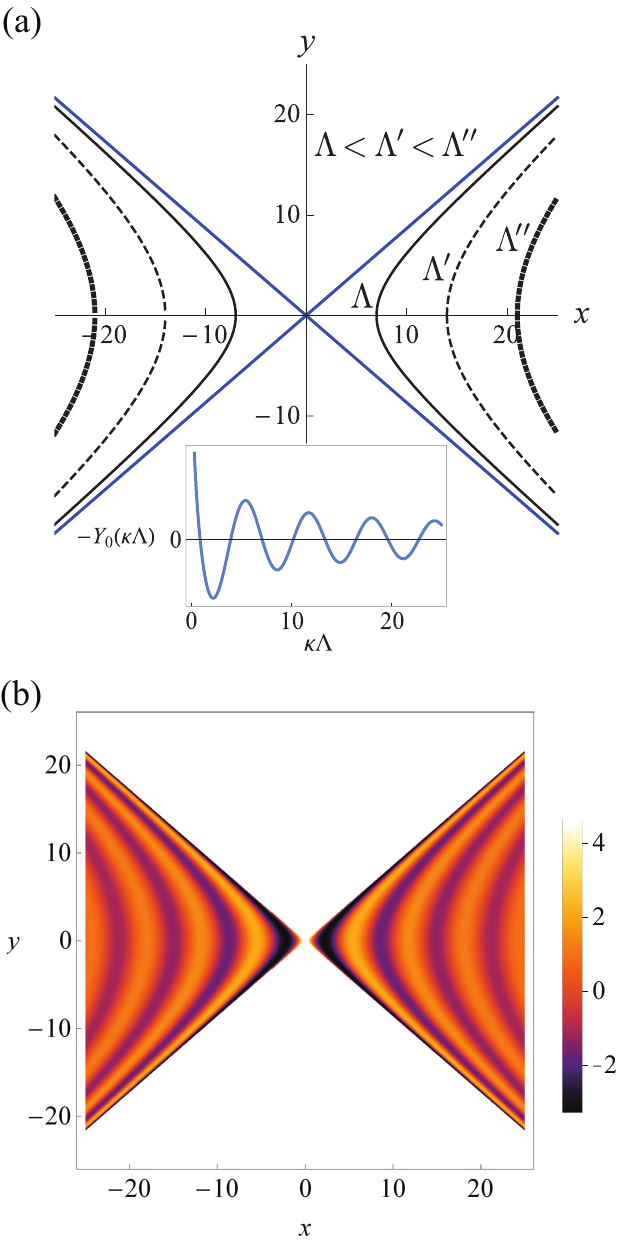}
\caption{Fourier transform of a hyperbola in the major region shows hyperbolic fringes. For the hyperbola in Fig.~\ref{Fig:regIII}a, this region contains the left and right quadrants of Fig.~\ref{Fig:regIII}b. The transform contours are given by a family of coasymptotic hyperbolas with three such hyperbolas shown in~(a) for different values of the hyper-radius $\Lambda, \Lambda'$ and $\Lambda''$. As the transform value shows oscillations in addition to a slow decay with the hyper-radius  (inset of~(a)), we have the resulting transform in this region~(b) exhibit hyperbolic fringes  decaying slowly from the separatrix lines.
}
\label{Fig:regI}
\end{figure}

Similarly to the case of the minor region, the transform value in the major region depends only on the hyper-radius $\Lambda$, and hence the transform contours are again given by the constant-$\Lambda$ curves. For our hyperbola in Fig.~\ref{Fig:regIII}, we obtain the contours from~(\ref{Eq:hyperadius}) as
\begin{equation}
\frac{x^2}{\Lambda^2/\epsilon_y} - \frac{y^2}{\Lambda^2/(-\epsilon_x)}  = 1
 \label{Eq:contourI}
\end{equation}
which also describes a family of hyperbolas parametrized by $\Lambda$ (see Fig.~\ref{Fig:regI}a), but this time, all with their major axes oriented along the $x$ direction. As before, the foci of these hyperbolas shift away linearly with an increasing hyper-radius, and their asymptotes are given by the same two straight lines of the separatrix. Now, contrary to before, the functional dependence of the transform on the hyper-radius, as illustrated in the inset of Fig.~\ref{Fig:regI}a, shows oscillations in addition to decay. This results in a fringe pattern in the major region with the undulations traced out by a series of coasymptotic hyperbolas~\cite{[{Readers familiar with special relativity would notice that such coasymptotoic hyperbolas also appear in describing the motion of a uniformly accelerated body in the spacetime diagram by Rindler coordinates. For details, see }]misner1973gravitation} (see Fig.~\ref{Fig:regI}b). Also note how the transform decays monotonically from the separatrix ridges.\\

\begin{figure*}[tb]
\centering
   \includegraphics[width=\textwidth]{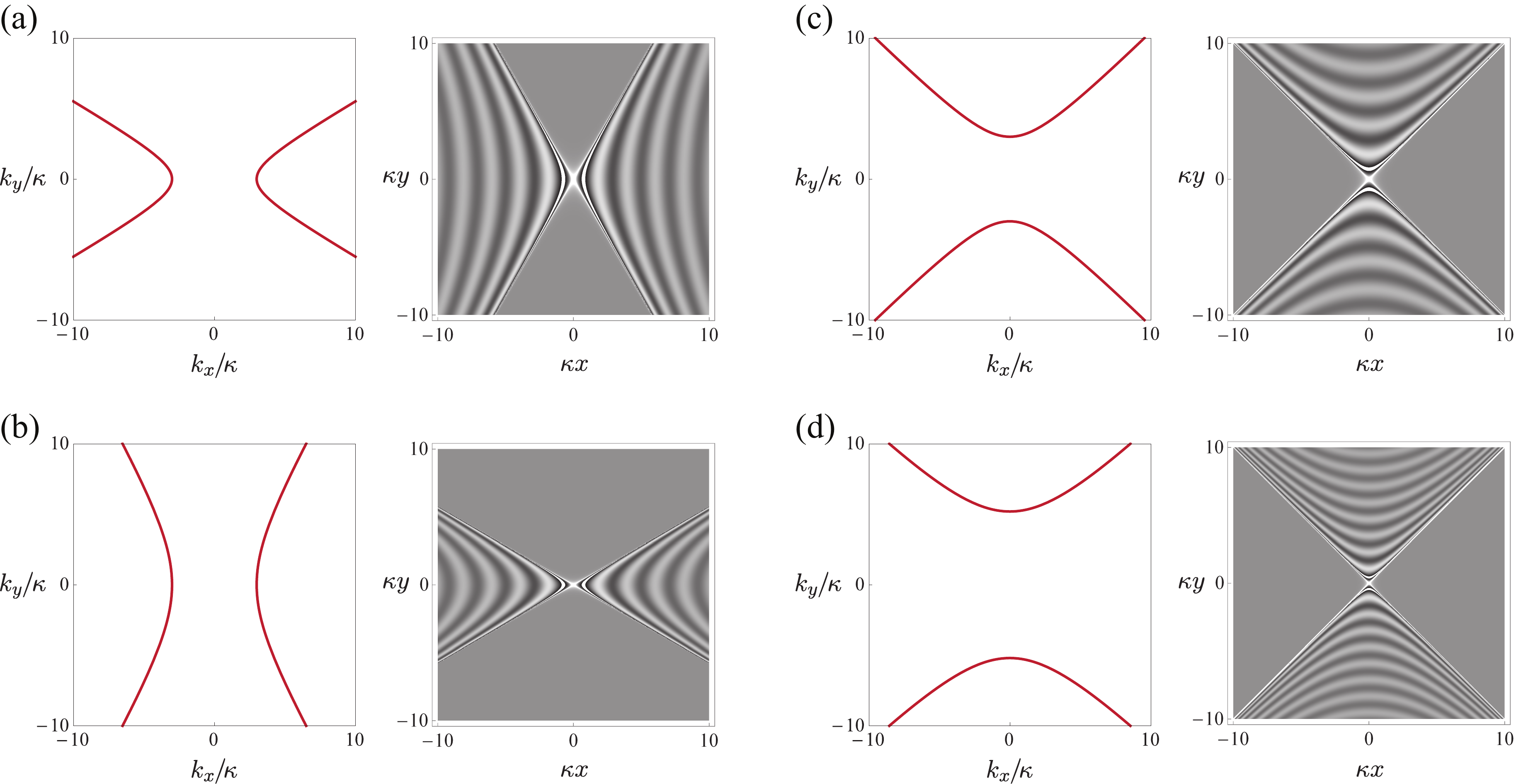}
\caption{Variations in eccentricity, orientation and overall size of the dispersion hyperbola (left panels) modify the transform pattern (right panels).
For hyperbolas described by~(\ref{Eq:hyperbola}), here we considered (a)~$-3\epsilon_x =  \epsilon_y =9$, (b)~$-\tfrac{1}{3}\epsilon_x =  \epsilon_y = 9$, (c)~$\epsilon_x =  -\epsilon_y =9$, and (d)~$\epsilon_x =  -\epsilon_y =27$. In all cases, the exponential profile in the minor region is overshadowed by the much larger transform amplitudes in the major region.
}
\label{Fig:other}
\end{figure*}

Combining the individual contributions of all three regions gives us the complete graphical representation of the Fourier transform of a hyperbola, akin to Fig.~\ref{Fig:intro}b, corresponding to the  fields radiated by a point source in a hyperbolic medium. Note the coherence among the three regions with a blow-up in the separatrix and decay on either side. Interestingly, the decay in the major region is much slower than that in the minor region. This is due to the fact that, away from the origin, the transform value in the major region scales as $ 1/{\sqrt{\kappa \Lambda}}$, whereas in the minor region it scales as  ${e^{- \kappa \Lambda}}/{\sqrt{\kappa \Lambda}}$~\cite{arfken2011mathematical}. As the dispersion hyperbolas change in eccentricity, orientation or size, the corresponding transform patterns change accordingly, as shown in Fig.~\ref{Fig:other}.
For hyperbolas with smaller eccentricity, the major region expands (see Fig.~\ref{Fig:other}a), whereas for larger eccentricity it contracts (see Fig.~\ref{Fig:other}b). Interchanging the directions of major and minor axes likewise swaps the corresponding major and minor regions in the transform (see Fig.~\ref{Fig:other}c). When the overall size of the hyperbola is scaled up, the fringes become more closely spaced (see Fig.~\ref{Fig:other}d). Now, why is the Fourier transform of a hyperbola the way it is, why does it have fringes on one region while in the other region it is structureless, how do we interpret it physically -- we answer these questions in the next section.

\section{Physical Interpretation}
\label{Sec:PI}
\begin{figure}[tb]
\centering
   \includegraphics[width=\columnwidth]{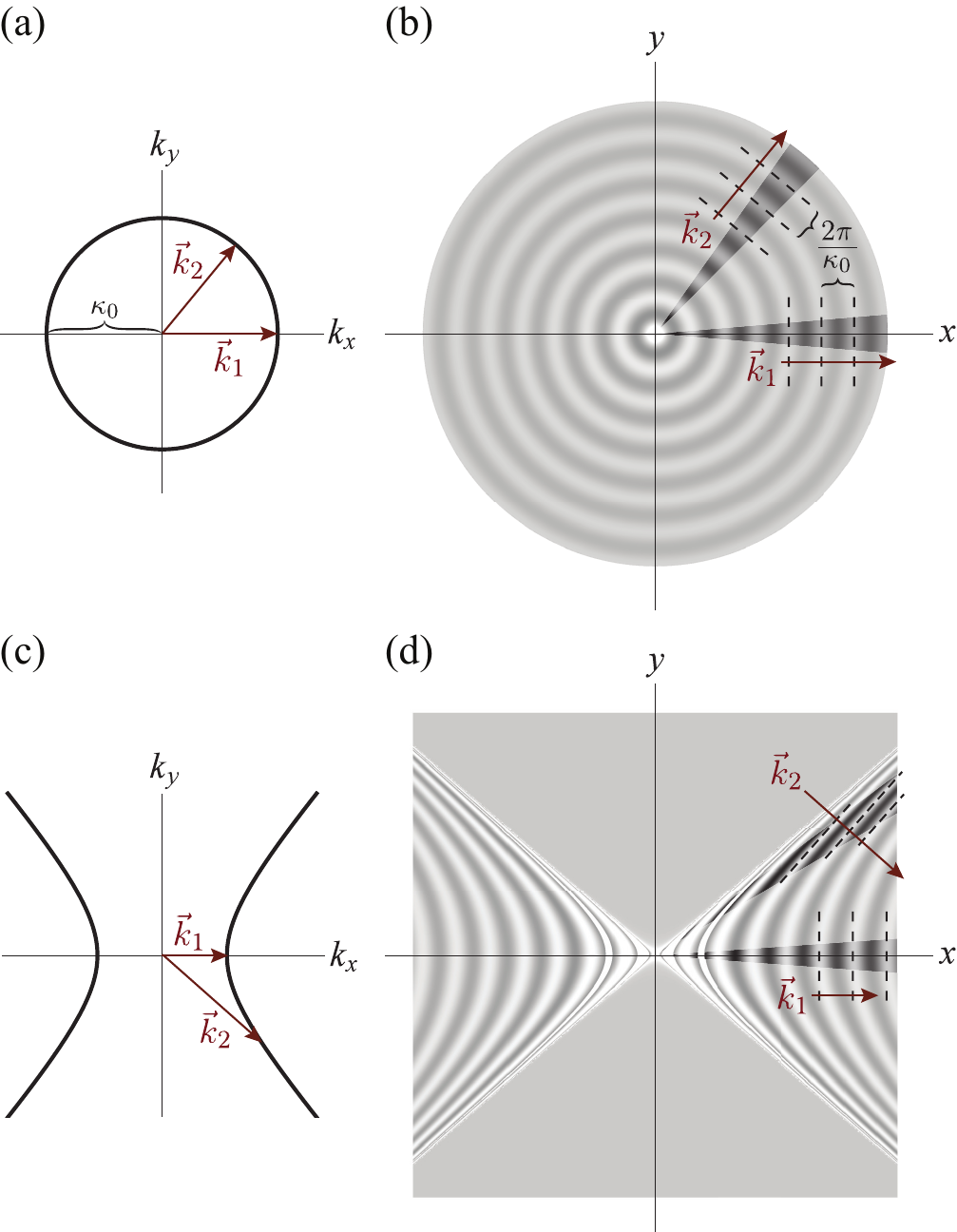}
\caption{With the  transform pairs shown for a circle~(a)--(b) and a hyperbola~(c)--(d), the respective Fourier patterns can be interpreted by the ``pitch--wavevector correspondence" which states that a  narrow sector (highlighted slices) in the real space, with its fringes approximated by a series of  parallel tangents (dashed lines) accommodating a certain pitch, provides us with a wavevector (red arrows) of the right magnitude and right direction, so that it corresponds to a valid momentum state in the reciprocal space. For the purpose of illustration, we have not shown the wavevectors in the opposite direction coming from the same sector. }
\label{Fig:PI}
\end{figure}

We start   by considering the simple case of a circular (isotropic) dispersion, and see how we can intuitively understand its Fourier pattern. As discussed in Section~\ref{Sec:M}, for a circle in reciprocal space with given radius $\kappa_0$ (see Fig.~\ref{Fig:PI}a), its inverse Fourier transform varies as $J_0(\kappa_0 r)$ (see Eq.~(\ref{Eq:circle})), and hence results in the familiar bullseye pattern~\cite{lezec2002beaming} in real space (see Fig.~\ref{Fig:PI}b). Owing to the asymptotic nature of Bessel functions~\cite{arfken2011mathematical}, the spatial variation of the transform away from the origin is well-described by a periodic pattern of  concentric circular fringes with periodicity ${2\pi}/{\kappa_0}$.\\

If we now consider a sector of the Fourier pattern with small angular width, the periodic arcs inside the sector would  tend to appear as a series of linear fringes. These fringes, paraxially mimicking a plane-wave segment,   correspond to a wavevector (spatial harmonic component) whose magnitude is given by the pitch of the fringes and whose direction is given along the direction of the periodic variation (normal to the wavefront). One can get a visual sense of the linear fringes for a given sector by drawing parallel tangent lines on the circular arcs, as shown in Fig.~\ref{Fig:PI}b. \\

Since one can choose a sector in any direction, one can obtain a wavevector that also changes in direction but does not change in magnitude. By combining all such wavevector contributions we can recover back the circle in reciprocal space of Fig.~\ref{Fig:PI}a. This correspondence between the pitch of different sectors of the real space and the wavevector or momentum states of the reciprocal space provides us with a nice physical interpretation of the Fourier transform of a circle. This interpretation naturally extends also to the case of weak anisotropy, for which the iso-frequency contour is an ellipse.\\

To see if the pitch--wavevector correspondence can also explain the Fourier pattern of a hyperbola, let us consider, without loss of generality, the hyperbolic dispersion shown in Fig.~\ref{Fig:PI}c described by~(\ref{Eq:hyperbola}) with $\epsilon_x<0$ and $\epsilon_y>0$. As discussed in Section~\ref{Sec:GR}, we note that each of the hyperbolic fringes in the major region of the transform (see Fig.~\ref{Fig:PI}d) is characterized by a unique value of the hyper-radius $\Lambda$. Moreover, away from the origin, due to the asymptotic behavior of the modified Bessel functions~\cite{arfken2011mathematical}, two neighboring fringes differ in their hyper-radii by $\Delta \Lambda = 2 \pi/{\kappa}$.\\

Now, if we consider a narrow sector in the major region near the $x$ axis (see Fig.~\ref{Fig:PI}d), we can approximate the hyperbolic arcs inside the sector by a set of linear fringes which are oriented along the $y$ direction. Since $y = 0$ mostly within the sector, we have the corresponding pitch from~(\ref{Eq:hyperadius}) as $\Delta x = {\Delta \Lambda}/{\sqrt{\epsilon_y}} $. As a result, this sector contributes a wavevector along the $k_x$ direction with magnitude $2 \pi/{\Delta x}$ which coincides with the apex of the hyperbola $(\sqrt{\epsilon_y}\, \kappa,0)$ in the reciprocal space (see Fig.~\ref{Fig:PI}c).\\

Next we consider a sector near the asymptotes of the hyperbolic fringes. It is clearly evident from Fig.~\ref{Fig:PI}d that as the sector moves closer to one of the asymptotic lines of the  separatrix, the fringes become denser and parallel to the asymptote. As a result the corresponding wavevector progressively becomes larger in magnitude and perpendicular to the separatrix line, yielding a momentum state lying on the asymptote of our original hyperbola in the reciprocal space (see Fig.~\ref{Fig:PI}c).\\

With the pitch--wavevector correspondence accounting for the momentum states at the apex and the asymptote -- two extreme cases with highest and lowest curvature on the hyperbola -- we turn to see if the interpretation also holds  for any intermediate case. Indeed it turns out, as we show below, that a sector anywhere in the bulk of the major region,  with its fringes approximated by a series of  parallel tangent lines, provides us with a wavevector  of the right magnitude and right direction, so that it corresponds to a momentum state lying on the original hyperbola in the reciprocal space.\\

\begin{figure}[tb]
\centering
   \includegraphics[width=\columnwidth]{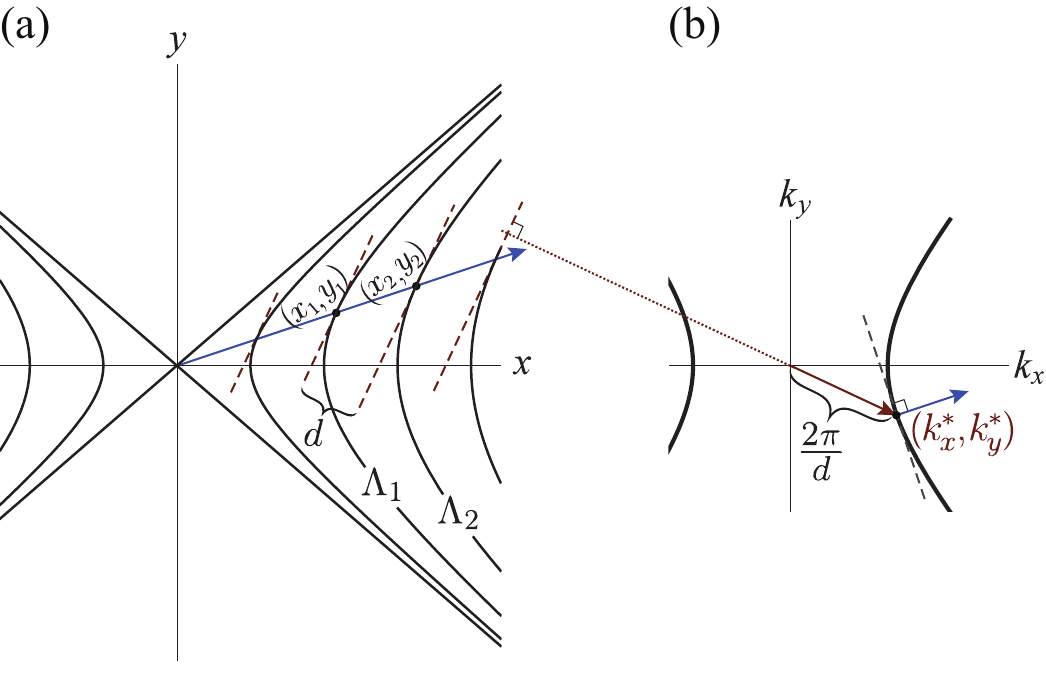}
\caption{The pitch--wavevector correspondence can be proved in the following steps by showing (i)~if the angular direction of a given sector of the Fourier pattern in real space~(a) is  described by a ray passing through the origin~(blue arrow), then the points where the ray intersects the contours of the hyperbolic fringes accommodate  tangents~(dashed lines) which are all parallel; (ii)~the separation distance $d$ between any two neighboring tangents is constant across all fringes -- leading to the formation of a well-defined grating pattern; and finally (iii)~the direction of the wavevector corresponding to this pattern~(red arrow) intersects the transform hyperbola in the reciprocal space~(b) at a distance $2\pi/d$ away from the origin. Additionally, we have the direction of the transform gradient (up to a sign) given by the same ray describing the angular direction of the sector.}
\label{Fig:proof}
\end{figure}

To establish the correspondence concretely, let us consider two neighboring fringes in the major region characterized by hyper-radii $\Lambda_1$ and $\Lambda_2$ such that $\Lambda_2-\Lambda_1 = 2\pi/\kappa$ (see Fig.~\ref{Fig:proof}a). Now in order for a straight line with slope $m$ to be tangent on either of these fringes, it has to be steeper than the fringe asymptotes, that is
\begin{equation}
|m| > \sqrt{-\frac{\epsilon_y}{\epsilon_x}}.
 \label{Eq:slopecon}
\end{equation}
If two such parallel straight lines with the same slope $m$ touch the hyperbolic fringes on points $(x_1,y_1)$ and $(x_2,y_2)$ so that from~(\ref{Eq:hyperadius}) we have
\begin{equation}
\epsilon_y x_i^2  +\epsilon_x y_i^2    = \Lambda_i^2  \quad \text{for } i=1,2
 \label{Eq:hyppoint}
\end{equation}
then the equations describing these two tangent lines are given by
\begin{equation}
\epsilon_y x_i\, x  +\epsilon_x y_i \, y    = \Lambda_i^2  \quad \text{for } i=1,2.
 \label{Eq:tangent}
\end{equation}
Equating the slopes gives us
\begin{equation}
\frac{1}{m} \left(- \frac{\epsilon_y}{\epsilon_x} \right) = \frac{y_1}{x_1} = \frac{y_2}{x_2} 
 \label{Eq:slope}
\end{equation}
which suggests that tangent lines with a given slope touch the fringes at points which all lie in a straight line passing through the origin (see Fig.~\ref{Fig:proof}a). This allows us to approximate the fringes within a given sector by a series of tangent lines which are all parallel to one another.\\

Now letting $y_2/y_1 = x_2/x_1 = \rho$ we obtain from~(\ref{Eq:hyppoint})  $\Lambda_2 = \rho \Lambda_1$ and from~(\ref{Eq:tangent}) the two tangents described as $\epsilon_y x_1\, x  +\epsilon_x y_1 \, y    = \{ \Lambda_1^2, \rho\Lambda_1^2\}$, which admit a separation distance 
\begin{equation}
d = \frac{2\pi}{\kappa}\sqrt{\frac{ \frac{1}{\epsilon_y} m^2  +  \frac{1}{\epsilon_x} }{m^2+1}}.
 \label{Eq:separation}
\end{equation}
Here the positivity of the quantity under square root follows from~(\ref{Eq:slopecon}). Since the distance does not depend on the individual values of the two hyper-radii, any pair of neighboring tangent lines has the same separation distance across all fringes. This allows us to treat a sector of the hyperbolic fringes as a linear grating pattern with a constant pitch.\\

As the direction of the wavevector corresponding to this grating pattern is normal to the tangents, it will point along a line with slope $-\sfrac{1}{m}$ in the reciprocal space. Let us consider the momentum state $\vec{k^*} = (k_x^*,k_y^*)$ on our hyperbola in the reciprocal space which lies along that direction (see Fig.~\ref{Fig:proof}b). Then
\begin{equation}
\begin{aligned}
 k_y^* = - \frac{1}{m}k_x^* \quad \text{and}\\
 \frac{(k_x^*)^2}{\epsilon_y}  +\frac{(k_y^*)^2}{\epsilon_x}=\kappa^2
\end{aligned}
\label{Eq:reci}
\end{equation}
which upon solving for the magnitude $\|\vec{k^*}\| = \sqrt{k_x^2+k_y^2}$ gives
\begin{equation}
\|\vec{k^*}\|= 2\pi/d.
 \label{Eq:final}
\end{equation}
This implies that the magnitude of the momentum state also matches the wavevector corresponding to the grating pattern. In other words, parallel tangents of the hyperbolic fringes in real space contribute a wavevector state that lies on the hyperbola in the reciprocal space -- establishing the pitch--wavevector correspondence for any arbitrary sector.\\

It is interesting to note one additional correspondence pertaining to the angular direction of the sector. The tangent line on our hyperbola in the reciprocal space at point $(k_x^*,k_y^*)$ can be described as 
\begin{equation}
\frac{k_x^*}{\epsilon_y}  \, k_x  +\frac{k_y^*}{\epsilon_x}  \, k_y     = \kappa^2 
 \label{Eq:tangentF}
\end{equation}
admitting a slope 
\begin{equation}
m' = \frac{\epsilon_x}{\epsilon_y}\left(- \frac{k_x^*}{k_y^*} \right) = \frac{\epsilon_x}{\epsilon_y} m
 \label{Eq:tangentFslope}
\end{equation}
with the last equality following from~(\ref{Eq:reci}). This in turn gives us the slope of the normal  (see Fig.~\ref{Fig:proof}b) that is perpendicular to the tangent at $(k_x^*,k_y^*)$ as $-\frac{1}{m'} = \frac{1}{m} (- \frac{\epsilon_y}{\epsilon_x} )$ which, upon comparing with~(\ref{Eq:slope}), matches exactly with the slope of the line connecting all the touching points on the hyperbolic fringes in real space. This implies the angular direction of a narrow sector in real space corresponds with the direction of the transform gradient $\nabla_{\vec{k}}\: F(k_x^*,k_y^*)$ in the reciprocal space (see Eq.~(\ref{Eq:hypFD})) evaluated at the wavevector $(k_x^*,k_y^*)$ that  the sector contributes.\\

This additional correspondence gives us the direction of group velocity of the wave for each sector of the field profile, and affirms that the energy indeed propagates away from the source  in a radially outward manner -- a fact not obvious for a field distribution that is not circularly symmetric as a bullseye pattern.\\

With the pitch-wavevector correspondence now on firm footing~\cite{pstheorem},  we can finally interpret the Fourier pattern of a hyperbola in the major region as a coherent, smoothly-connected ensemble (atlas) of linear fringes, each of which  locally contributes (charts) a wavevector state of the original hyperbola in the reciprocal space. This is quite similar to how a crystal lattice Bragg-scatters an incoming X-ray radiation by different amounts through its different crystal planes~\cite{ashcroft2022solid}. Meanwhile, in the minor region,  the absence of any structure implies that this region can not contribute any momentum state. In the context of the hyperbolic wave excitation, this simply translates to the evanescent field decay from the excitation source at the origin, since there are no propagation states along these directions.\\

\section{Huygens' principle in hyperbolic media}
\label{Sec:D}
Huygens' principle states that every point on a wavefront serves as a secondary source of spherical wavelets, and the new wavefront at a later instant is given by the envelope of all such secondary waves~\cite{huygens1690treatise, *[{English translation by S. P. Thompson: }]huygens1912treatise, baker2003mathematical}. Being the first successful theory of wave propagation, Huygens' principle can account for waves of various shapes including spherical, planar, cylindrical, and even elliptical~\cite{bergstein1966huygens}. Here we show this principle can also be extended for waves with hyperbolic wavefronts.\\

Hyperbolicity of the wavefronts arises in acoustics~\cite{li2009experimental, shen2015broadband, quan2019hyperbolic}, elastodynamics~\cite{oh2014truly},  magnonics~\cite{macedo2019engineering} and predominantly in optics~\cite{jacob2006optical,salandrino2006far} with realizations well-studied for bulk electromagnetic waves in hyperbolic metamaterials~\cite{liu2007far, ma2018experimental} and surface electromagnetic waves such as hyperbolic phonon polaritons in low symmetry crystals~\cite{ma2018plane, ma2021ghost}. In general, if the anisotropy of the supporting medium is so extreme that material responses in two orthogonal directions are of opposite nature, for example, metallic and dielectric, and are characterized by susceptibility measures of opposite signs, then the wave dispersion  assumes a hyperbolic shape~\cite{jacob2006optical,salandrino2006far}. As a result, a point source that excites all the states on the hyperbolic dispersion would emit waves with hyperbolic wavefronts.\\

\begin{figure}[tb]
\centering
   \includegraphics[width=\columnwidth]{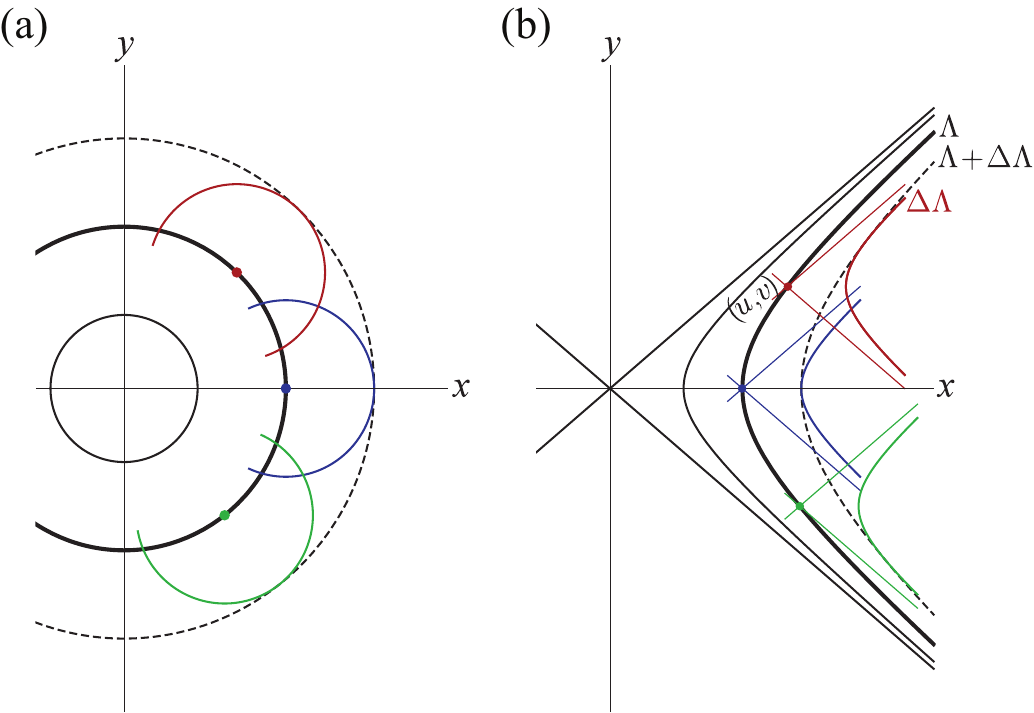}
\caption{Huygens' principle for wave propagation can be generalized to include hyperbolic wavefronts. For the traditional isotropic case~(a) where a point source emits circular waves, all points on the primary wavefront~(solid black) can be thought of as secondary sources of circular wavelets~(color), and their envelope~(dashed) gives the new wavefront at a later instant. For a similar point excitation inside a hyperbolic medium~(b), the primary wavefront, the secondary wavelets and their envelope conform to Huygens' principle, although here they assume hyperbolic shape given by the Fourier transform of the dispersion curve.
}
\label{Fig:point}
\end{figure}

In order to extend Huygens' principle to hyperbolic wavefronts, let us consider point source excitation in both a isotropic dielectric and a hyperbolic media, as shown in Fig.~\ref{Fig:point}. As well-established for isotropic medium (see Fig.~\ref{Fig:point}a), the primary wavefront, the secondary wavelets, and their envelope -- all are circular in shape, and Huygens' principle follows in a straightforward manner. For the hyperbolic medium (see Fig.~\ref{Fig:point}b), we observe that the primary wavefront and the secondary wavelets are given by the Fourier transform of the hyperbolic dispersion curve, and hence, they assume hyperbolic shape. Huygens' principle will extend to this case if the envelope touching all the secondary wavelets is shown to be hyperbolic as well -- which we accomplish below.\\

Let us describe the dispersion of our hyperbolic medium by~(\ref{Eq:hyperbola}) with $\kappa=\omega/c$ where $\omega$ is the excitation frequency of the source placed at the origin, and $c$ is the speed of light (see Fig.~\ref{Fig:point}b for the case of $\epsilon_x<0$ and $\epsilon_y>0$). Also, let us assume our primary wavefront originates at time $t=0$, and at a later time $t=t$, attains a position characterized by a hyperbola of hyper-radius $\Lambda = c\,t$. Then any point $(u,v)$ on this wavefront will satisfy 
\begin{equation}
\epsilon_y u^2  +\epsilon_x v^2    = \Lambda^2  
 \label{Eq:uv}
\end{equation}
and hence can be parametrized by a hyperbolic angle $\tau$ as
\begin{equation}
u(\tau)= \frac{\Lambda}{\sqrt{\epsilon_y}} \cosh{\tau}      \quad \text{and} \quad v(\tau) =\frac{\Lambda}{\sqrt{-\epsilon_x}} \sinh{\tau}.
 \label{Eq:tau}
\end{equation}
Now according to Huygens' principle each of these points will act as a secondary source, with three such sources shown in Fig.~\ref{Fig:point}b. Then at  $\Delta t$ time later, the secondary wavelets emanating from these sources will be characterized by a hyper-radius $\Delta\Lambda = c \Delta t$. If we let
\begin{equation}
W(x,y;\tau) = \epsilon_y \left(x-u(\tau)\right)^2  +\epsilon_x (y-v(\tau))^2    - (\Delta\Lambda)^2
 \label{Eq:family}
\end{equation}
describe  the family of all hyperbolic wavelets, then their envelope~\cite{bruce1992curves} is obtained by the simultaneous solution of 
\begin{equation}
W(x,y;\tau) = 0  \quad \text{and} \quad \frac{\partial W}{\partial\tau} = 0.   
 \label{Eq:W}
\end{equation}
Upon solving we  have the envelope specified as
\begin{equation}
x= \frac{\Lambda+\Delta\Lambda}{\sqrt{\epsilon_y}} \cosh{\tau}      \quad \text{and} \quad y =\frac{\Lambda+\Delta\Lambda}{\sqrt{-\epsilon_x}} \sinh{\tau}
 \label{Eq:envelope}
\end{equation}
which also admits a shape of a hyperbola given by
\begin{equation}
\epsilon_y x^2  +\epsilon_x y^2    = (\Lambda+\Delta\Lambda)^2.  
 \label{Eq:new}
\end{equation}
Since $\Lambda+\Delta\Lambda = c(t+\Delta t)$, we have the new wavefront at time $t+\Delta t $ given exactly by the above envelope of the secondary wavelets, proving Huygens principle for hyperbolic wavefronts. Also note, by comparing~(\ref{Eq:tau})~and~(\ref{Eq:envelope}), we have $x/y = u/v $, which implies that the direction from a secondary source to the point where the envelope touches its corresponding wavelet is the same as that of energy propagation of the primary wave.\\

\begin{figure}[tb]
\centering
   \includegraphics[width=\columnwidth]{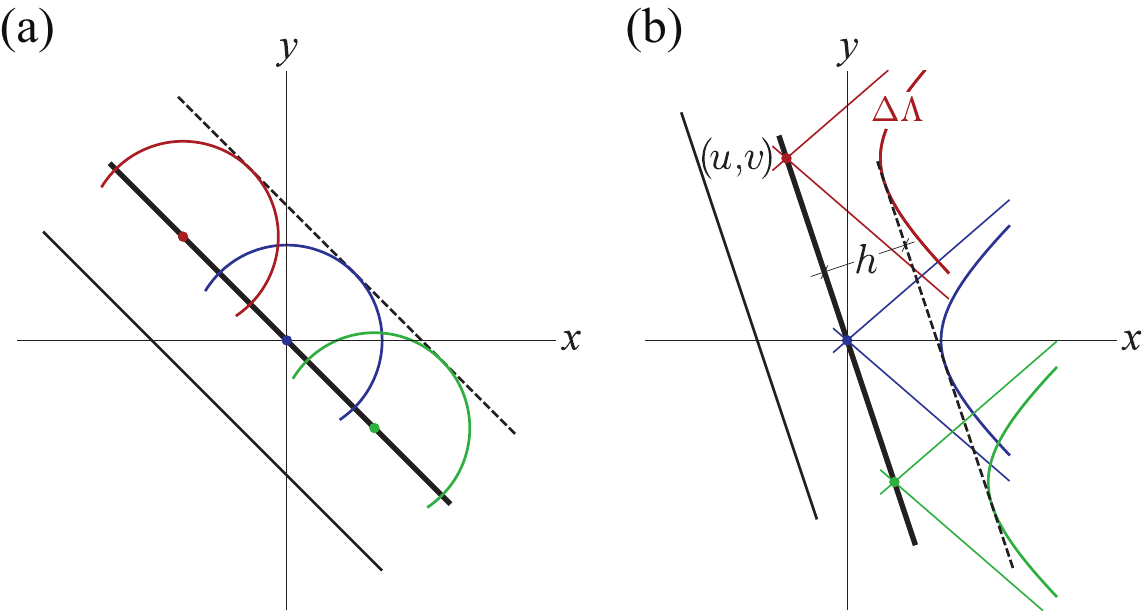}
\caption{Huygens' principle is demonstrated for planar waves inside an isotropic dielectric~(a) and a hyperbolic medium~(b). In both instances, the direction from secondary source to the touching point on the envelope gives the direction of energy flow. Note how Huygens' picture for the hyperbolic case naturally allows for misalignment between the propagation wavevector and the energy-flow (Poynting) vector.
}
\label{Fig:rec}
\end{figure}

Huygens' principle can also be extended to planar wavefronts propagating inside a hyperbolic medium. Similar to plane waves in a isotropic dielectric (see Fig.~\ref{Fig:rec}a), Huygens' principle retains the rectilinear structure of planar wavefronts in a hyperbolic medium (see Fig.~\ref{Fig:rec}b for the case of $\epsilon_x<0$ and $\epsilon_y>0$). We establish this fact below paralleling the proof given for hyperbolic wavefront.\\

Let us describe our plane wave of slope $m$ with a wavevector $\vec{k^*}$, and place our coordinate origin on the primary wavefront so that any of its point $(u,v)$ can be parametrized as 
\begin{equation}
u(\tau)= \tau      \quad \text{and} \quad v(\tau) = m \tau.
 \label{Eq:tau2}
\end{equation}
As before, these points will act as secondary sources of hyperbolic wavelets, each of which can be characterized by a hyper-radius $\Delta\Lambda = c \Delta t$ after an elapsed time $\Delta t$. With $W(x,y;\tau)$ defined in~(\ref{Eq:family}) describing the family of all such hyperbolic wavelets, ${\partial W}/{\partial\tau} = 0$ yields
\begin{equation}
\frac{y-v}{x-u} = \frac{1}{m}\left(-\frac{\epsilon_y}{\epsilon_x}\right).
 \label{Eq:W2}
\end{equation}
Substituting this in the other condition $W(x,y;\tau)=0$ gives us the final envelope
\begin{equation}
y = m x + \Delta \Lambda\, \sqrt{ \frac{1}{\epsilon_y} m^2  +  \frac{1}{\epsilon_x} }
 \label{Eq:new2}
\end{equation}
which also describes a line with the same slope $m$. Its separation distance $h$ from the primary wavefront is given by
\begin{equation}
h = \Delta \Lambda\,\sqrt{\frac{ \frac{1}{\epsilon_y} m^2  +  \frac{1}{\epsilon_x} }{m^2+1}} = \Delta \Lambda\, \frac{\kappa}{\|\vec{k^*}\|} = \frac{\omega}{\|\vec{k^*}\|}\Delta t
 \label{Eq:shift}
\end{equation}
where we have made use of the relations~(\ref{Eq:final})~and~(\ref{Eq:separation}) in the intermediate step. Since ${\omega}/{\|\vec{k^*}\|}$ is the phase velocity of our plane wave, the envelope above gives us the new wavefront at $\Delta t$ time later. This validates Huygens' principle for planar wavefronts in hyperbolic medium. \\

Also, by comparing~(\ref{Eq:W2})~and~(\ref{Eq:slope}), we can reaffirm that the direction from secondary source to the touching point on the envelope aligns with the direction of energy flow. In contrast to the symmetric case of circular wavelets, this direction does not necessarily align with the direction of phase propagation. Note how Huygens' principle applied to hyperbolic wavelets leads to an intuitive understanding of this misalignment.\\

\begin{figure}[tb]
\centering
   \includegraphics[width=0.8\columnwidth]{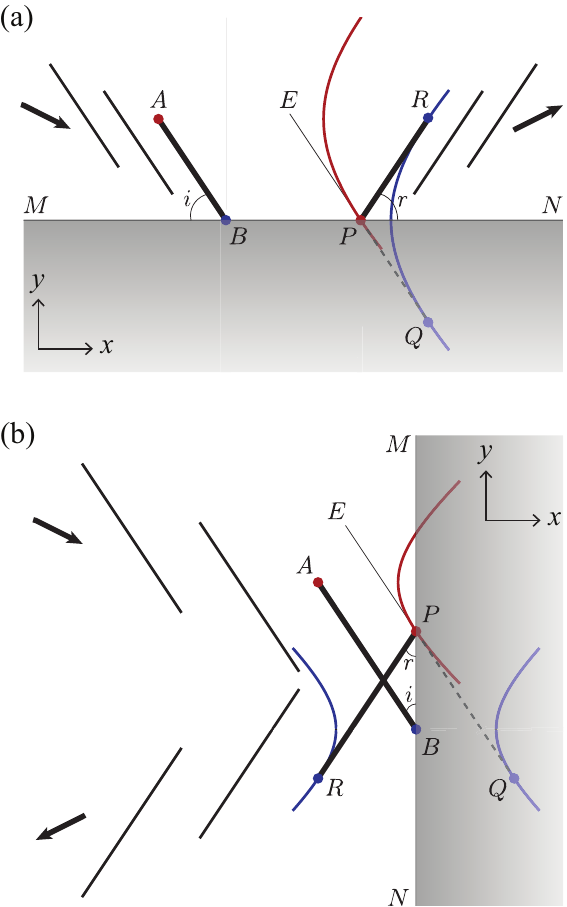}
\caption{Huygens' principle with hyperbolic wavelets can be used to prove the law of reflection $\angle i = \angle r$ when a plane wave is incident from a hyperbolic medium on an interfacial cut that runs parallel to the   major~(a) or minor~(b) axis of the dispersion hyperbola.
}
\label{Fig:reflection}
\end{figure}

With an understanding of wave propagation inside a hyperbolic medium based on Huygens' principle, we next show the same principle can also prove the law of reflection. Toward that end, let us consider a planar wavefront $AB$ incident from a hyperbolic medium on a practical interface $MN$ that runs parallel to the  major or minor axis of the dispersion hyperbola, as illustrated in Figs.~\ref{Fig:reflection}a~and~\ref{Fig:reflection}b, respectively. 
%Let the incident wavefront $AB$ make an angle $\angle i$ with the interface $MN$.
 If some time $\Delta t$ later, point $P$ on the interface becomes the new position for point $A$, then $P$ would lie on some hyperbolic wavelet of hyper-radius $\Delta \Lambda = c\Delta t$ originating from $A$. If one draws the corresponding wavelet for point $B$ with the same hyper-radius, then according to the foregoing Huygens' construction for plane wave propagation, tangent $PQ$ would have been the new wavefront -- had there been no reflecting medium. But due to reflection, the whole construction flips, and tangent $PR$ becomes the reflected wavefront, with the angle of reflection~$\angle r = \angle QPN \text{ (reflection symmetry) } =\text{ opposite } \angle EPM  =\text{ corresponding } \angle ABM = \text{ angle of incidence } \angle i $. Note how the argument presented only works for interfaces that are aligned with the symmetry axes of the dispersion hyperbola, and in general, we do not have $\angle i = \angle r$ for arbitrary orientation of the interfacial cut.\\

\begin{figure}[tb]
\centering
   \includegraphics[width=0.8\columnwidth]{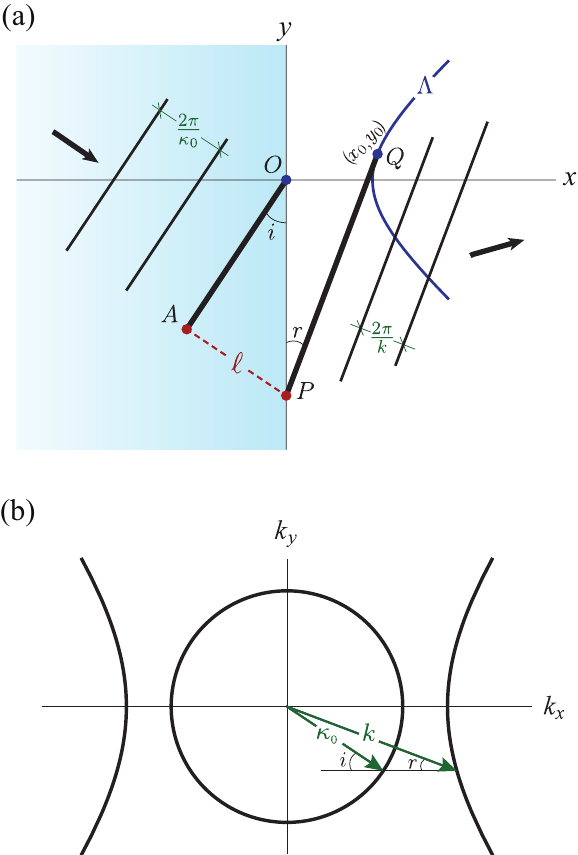}
\caption{Huygens' principle with hyperbolic wavelets can be used to prove the equivalent of Snell's law. (a)~When a plane wave refracts  into a hyperbolic medium from an isotropic dielectric, Huygens' construction gives $ \kappa_0 \sin i=k \sin r$, which implies in the associated reciprocal space~(b) that momentum component along the direction of interface is conserved.
}
\label{Fig:refraction}
\end{figure}

Next we show Huygens' principle with hyperbolic wavelets can be used to prove the equivalent of Snell's law. We accomplish this for the case shown in Fig.~\ref{Fig:refraction} where a plane wave refracts from an isotropic dielectric into a hyperbolic medium characterized by~(\ref{Eq:hyperbola})  with $\epsilon_x<0$ and $\epsilon_y>0$. As Fig.~\ref{Fig:refraction}a shows, we consider the incident wave to have a wavenumber $\kappa_0$ and its wavefront $OA$ to make an angle $\angle i$ at the origin with the refracting interface which runs along the $y$ axis. Let us assume it takes some time $\Delta t$ for the point $A$ to reach  the interface at point $P$, so that the distance traveled $\ell = \frac{\omega}{\kappa_0} \Delta t$ where $\omega$ is the frequency. For the same time duration, let us describe the corresponding hyperbolic wavelet emanating from the origin $O$ by
 \begin{equation}
\epsilon_y x^2  + \epsilon_x y^2    = \Lambda^2 
 \label{Eq:refr_hyp}
\end{equation}
with the hyper-radius $\Lambda = \frac{\omega}{\kappa}\Delta t$. Then the tangent $PQ$ drawn on this hyperbola would give us the refracted wavefront which we assume to make an angle $\angle r$ with the interface and constitute a refracted wave of wavenumber $k$ inside the hyperbolic medium. Now, if the coordinate of the touching point $Q$ is $(x_0, y_0)$, then the tangent line is described by
\begin{equation}
\epsilon_y x_0\, x  +\epsilon_x y_0 \, y    = \Lambda^2
 \label{Eq:refr_tangent}
\end{equation}
which admits a slope $m$ and intercept $s$, suppose. Since, the point $Q$ lies on the hyperbolic wavelet, its coordinate $(x_0. y_0)$ would satisfy~(\ref{Eq:refr_hyp}), which gives us
\begin{equation}
\frac{s^2}{\Lambda^2} =\frac{1}{\epsilon_y} m^2  +  \frac{1}{\epsilon_x} = (m^2+1)\frac{\kappa^2}{k^2}
 \label{Eq:refr_m}
\end{equation}
where the last step follows from~(\ref{Eq:separation})~and~(\ref{Eq:final}). Using the relations $m = {1}/{\tan r}$ and $s = \ell/\sin i$,~(\ref{Eq:refr_m}) finally yields
\begin{equation}
 \kappa_0 \sin i=k \sin r 
 \label{Eq:snell}
\end{equation}
which is the equivalent of Snell's law in this geometry. Figure~\ref{Fig:refraction}b shows the implication of this relation in the reciprocal space, which simply states that momentum along the  direction of  interface is conserved.\\

\begin{figure}[tb]
\centering
   \includegraphics[width=0.8\columnwidth]{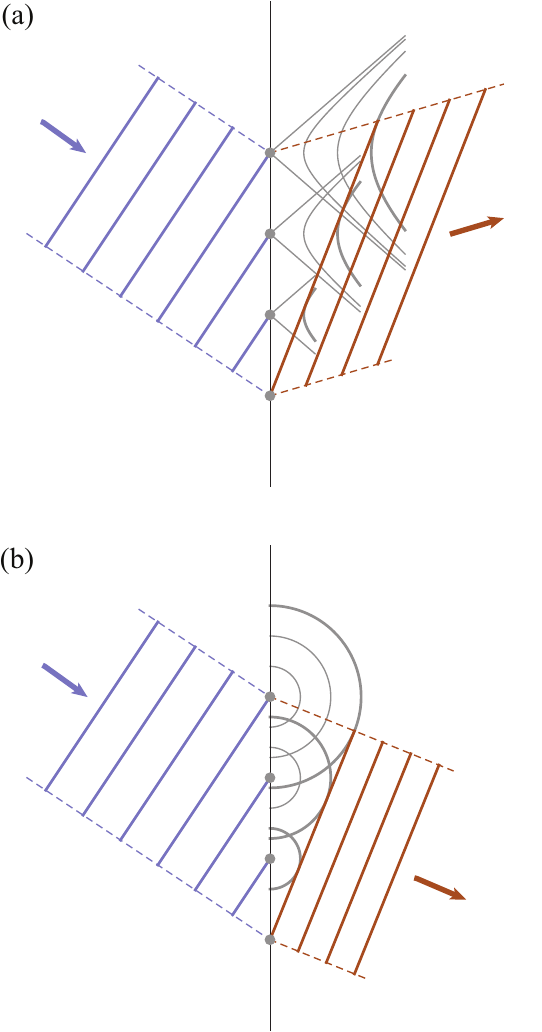}
\caption{Negative refraction follows naturally from Huygens' principle applied to hyperbolic wavelets. (a)~When light passes from an isotropic dielectric into a hyperbolic medium, then depending on the orientation of the interface, it can bend the ``wrong" way. This is to be contrasted with (b)~conventional refraction  involving two isotropic media.
}
\label{Fig:NegRef}
\end{figure}

Figure~\ref{Fig:refraction}a also illustrates how Huygens' principle applied to hyperbolic wavelets naturally leads to the concept of negative refraction~{\cite{Veselago_1968,smith2003electromagnetic, pendry2004negative, hoffman2007negative}. For its key importance in modern optics, we highlight this effect separately in Fig.~\ref{Fig:NegRef}. When light refracts into a hyperbolic medium from an isotropic dielectric through an interface along which the permittivity component of the hyperbolic material is positive, Huygens' construction, as shown in Fig.~\ref{Fig:NegRef}a, allows light to refract with a negative angle, and hence, energy flows the ``wrong" way along the tangential direction. This is to be contrasted with the conventional (positive) refraction picture in Fig.~\ref{Fig:NegRef}b involving two dielectrics with different refractive indices which was originally studied by Huygens~\cite{huygens1690treatise} through his circular wavelets.\\

\begin{figure}[tb]
\centering
   \includegraphics[width=0.75\columnwidth]{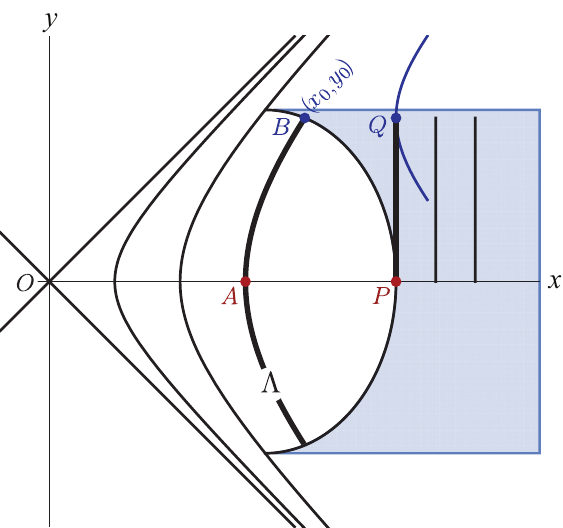}
\caption{Huygens' principle extended to hyperbolic wavelets can find applications in the task of lens construction. Here, the curved interface separating two hyperbolic media collimates a point source excitation from one side into a plane wave in the other.
}
\label{Fig:lens}
\end{figure}

Finally, to complete Huygens' picture for hyperbolic wavelets, we turn to a recent work on lens construction~\cite{bachmat2023lenses}. Figure~\ref{Fig:lens} shows such an arrangement where the curved interface between two hyperbolic media collimates a point source excitation into a plane wave. Here we assume the point source, placed at the origin $O$, emits in frequency $\omega$ inside the left hyperbolic medium whose dispersion is described simply by $k_x^2-k_y^2 = \kappa^2$. Let us assume, at one time instant, a hyperbolic wavefront $AB$ characterized by
\begin{equation}
 x^2- y^2 = \Lambda^2
 \label{Eq:wf}
\end{equation}
with $\Lambda = \overline{OA}$ being the hyper-radius, reaches the curved surface that interfaces with the right hyperbolic medium. If we choose the second medium to be $n>1$ times ``optically denser" than the first, that is, the dispersion of the right medium is characterized by $k_x^2-k_y^2 = (n\kappa)^2$, then the collimating interface needs to be semi-elliptical in shape  with an analytic form~\cite{bachmat2023lenses} given by, up to a scaling factor
\begin{equation}
 x^2(n^2-1) - \sqrt{2} n(n-1)x + \frac{1}{2}(n-1)^2 + y^2 = 0.
 \label{Eq:lens}
\end{equation}
Here, we seek to show Huygens' sources on this interface indeed produces a planar wavefront. Toward that end, let us assume it takes some time $\Delta t$ for point $A$ on the wavefront to reach the interface at point $P$ on the $x$ axis, so that the distance traveled $\overline{AP} = \frac{\omega}{\kappa}\Delta t$. At the same time duration, the secondary wavelet emanating from point $B$ would attain a hyper-radius $\overline{BQ} = \frac{\omega}{n\kappa}\Delta t$. Now it follows from~(\ref{Eq:lens}) that coordinate for point $P$ is $(\tfrac{1}{\sqrt{2}},0)$. If we let the coordinate for point $B$ to be $(x_0,y_0)$ then solving~(\ref{Eq:wf})~and~(\ref{Eq:lens}) simultaneously gives us
\begin{equation}
x_0 = \frac{1}{n}\left( \Lambda + \frac{n-1}{\sqrt{2}}\right).
 \label{Eq:x0}
\end{equation}
This in turn allows us to determine the $x$ coordinate for point $Q$ as $x_0 + \overline{BQ} = x_0 + \overline{AP}/n = x_0 + {\left(\frac{1}{\sqrt{2}} -\Lambda \right)}/{n} = \frac{1}{\sqrt{2}}$. Therefore, regardless of our choice of $\Lambda$, and the corresponding choice of secondary source on the interface, the apexes of all hyperbolic wavelets would lie on the $x=\frac{1}{\sqrt{2}}$ line at the same time instant when point $A$ reaches point $P$. This results in $PQ$ becoming the refracted wavefront which propagates as a plane wave in the denser hyperbolic medium. Lens construction methods based on a Minkowski metric~\cite{bachmat2023lenses} and hyperbolic wavelets are thus shown to be equivalent.\\

With Huygens' principle generalized for waves in hyperbolic media  by the Fourier transform of a hyperbola, we now have a   tool to explain negative refraction, wavevector -- Poynting-vector misalignment, lens geometry etc., without any need to refer to an auxiliary reciprocal space. This generalization also holds potential for applications in numerical simulations, complex lens designing, polaritonic manipulation, and wave field synthesis~\cite{berkhout1993acoustic}.

\section{Aliasing of the bullseye pattern}
\label{Sec:A}
If we place   a bullseye pattern~\cite{lezec2002beaming} on a digital screen, and keep zooming out so that the bullseye becomes progressively smaller and its concentric fringes progressively denser, then at one point the display resolution set by the finite pixel density of the screen will not be able to fully resolve the fine fringes, and one starts to see pattern in the bullseye that are not circular. In fact, one would start observing patterns that are very much similar to the hyperbolic fringes presented so far. \\

\begin{figure*}[tb]
\centering
   \includegraphics[width=\textwidth]{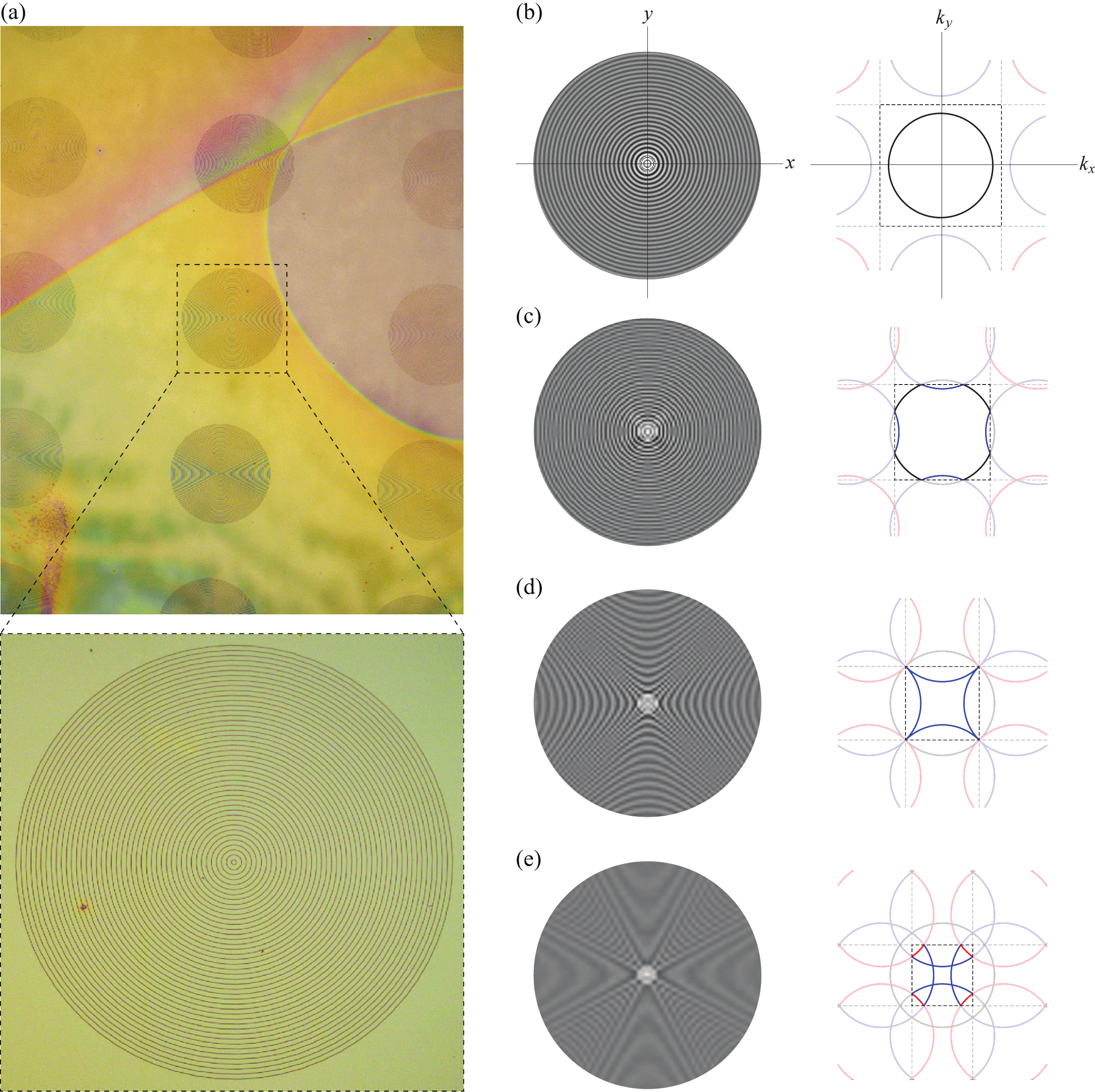}
\caption{``Hyperbolic fringes" are seen as aliasing artifacts when one degrades the resolution of an image of a bullseye pattern. (a)~Optical micrographs of germanium bullseyes written on a calcite substrate naturally show this phenomenon~(top); however, when zoomed in on an individual bullseye, the aliasing disappears~(bottom). The dashed black box marks a representative region for the zoomed-in view. (b)~When image sampling rate is high enough, no aliasing is present and the circle~(black) corresponding to the Fourier transform of the bullseye is within the first Brillouin zone~(transparent box). (c)~Hyperbolic fringes start appearing when image is downsampled beyond the Nyquist criterion, and circles~(blue) from nearest neighboring zones partially migrates in. (d)~Further downsampling results in the original circle~(black) lying completely outside the first Brillouin zone, and the aliased image is dominated by hyperbolic fringes along the horizontal and vertical directions. (e)~Even further downsampling allows the circles~(red) from the second nearest neighboring zones to migrate that give rise to hyperbolic fringes along the diagonal directions. In panels~(b)-(e), we have the bullseye image on the left and the corresponding Fourier space on the right.}
\label{Fig:aliasing}
\end{figure*}

Fig.~\ref{Fig:aliasing}a illustrates this aliasing effect using an array of germanium bullseye gratings deposited via electron-beam lithography on a calcite (100) surface, showing how optical microscope images obtained with different objective lenses can introduce aliasing. When a single bullseye is in focus, the circular pattern appears clean and aliasing-free; however, when a zoomed-out image with a larger field of view is taken, hyperbolic aliasing emerges. To explain this artifact, let us consider a bullseye pattern before aliasing took place, as shown in Fig.~\ref{Fig:aliasing}b. In the corresponding reciprocal space (shown in the right panel), the Fourier circle associated with the bullseye will be well-inside the first Brillouin zone formed by the square lattice of pixels. Owing to the periodic sampling process, there will be replicas of this circle in the neighboring Brillouin zones as well.\\

As one degrades the resolution of the bullseye image through downsampling, the spatial sampling rate and the size of the Brillouin zone will get smaller. At one point,  Nyquist criterion~\cite{soliman1990continuous} will no longer be satisfied, and we arrive at a situation (see Fig.~\ref{Fig:aliasing}c) where the Fourier circle is not fully contained within the first Brillouin zone. In this case the circles from the nearest neighboring zones partially migrate into the first Brillouin zone (analogous to band folding). The migrated circular arcs, which are similar in shape with hyperbolic arcs, give rise to the hyperbolic fringes in the bullseye image in the form of aliasing.\\

If one continues to downsample the image, then at one point the original Fourier circle will lie completely outside the first Brillouin zone. This is shown in Fig.~\ref{Fig:aliasing}d where there is no circular fringes visible in the downsampled image. Instead we see dominant presence of hyperbolic fringes in the horizontal and vertical directions owing to the migrated arcs from the nearest neighboring zones. If one downsamples the image further, circles from the second nearest neighboring zones starts migrating, and we see additional hyperbolic fringes in the diagonal directions (see Fig.~\ref{Fig:aliasing}e).\\

This image aliasing artifact illustrates how the Fourier transform of a hyperbola can be demonstrated in a simple setup. Admittedly, the Fourier pattern of a concave pair of circular arcs is not exactly the same as that of a truncated hyperbola, however, owing to their topological similarity for small arclength, the aliasing artifact qualitatively resembles the Fourier pattern of a hyperbola within a reasonable degree. Also note that the zooming-out process discussed at the beginning (Fig.~\ref{Fig:aliasing}a) shows the same aliasing effect as the downsampling process (Fig.~\ref{Fig:aliasing}b-e), as both these processes play with the relative size of the Fourier circle and the Brillouin zone. This allows one to quickly estimate the periodicity of a pixel array and the direction of sampling for a screen with an unknown resolution and orientation.

\vspace{-.2cm}

\section{Conclusion}
\label{Sec:C}

In this work, we have derived analytical expressions for the two-dimensional Fourier transform of the hyperbola, which leads to a closed-form expression for the radiation from a localized emitter in a hyperbolic medium, yielding rich physical interpretations of the radiation in such media with extreme anisotropy. If the given hyperbola has a finite extent, for instance limited by loss or nonlocality~\cite{yeh2005optical}, then the truncation of the support can be modeled by an appropriate two-dimensional window function. This would modify the Fourier pattern through a convolution by the corresponding blur kernel. As a result, areas on the pattern where the fringes were densely packed would now be too smeared to provide with the large wavevector states that lie outside the finite support of the hyperbola. Furthermore, there are physical examples where the hyperbolas are modified by a “shear” effect~\cite{passler2022hyperbolic}, with two of its diagonal branches being absent. The resulting Fourier transform reflect this distortion through the omission of fringes which are responsible for the missing wavevectors. With Huygens' principle generalized for waves in hyperbolic media, we now have a new mental tool to explain negative refraction, wavevector -- Poynting-vector misalignment, lens geometry, etc., without referring to the reciprocal space. We believe that this analysis opens interesting opportunities for the analysis and application of hyperbolic materials and metamaterials, and the design of nanophotonic devices based on these media.\\

\vspace{-.25cm}

The authors thank Viktoriia Rutckaia and J. Ryan Nolen for providing the images used in Figs.~\ref{Fig:intro}a~and~\ref{Fig:aliasing}a, respectively, and acknowledge support for this work from the Simons Foundation.

\end{document}